\documentclass[12pt]{article}
\usepackage{graphicx}
\usepackage{epsfig}
\usepackage{rotating}
\usepackage{dcolumn}
\usepackage{bm}
\usepackage{cite}%
\newcommand{\comment}[1]{}
\newcommand{\nn}{\nonumber}
\newcommand{\beq}{\begin{equation}}
\newcommand{\eeq}{\end{equation}}
\newcommand{\bea}{\begin{eqnarray}}
\newcommand{\eea}{\end{eqnarray}}
\newcommand{\barr}{\begin{array}}
\newcommand{\earr}{\end{array}}

\newcommand{\gsim}
{{\;\raise0.3ex\hbox{$>$\kern-0.75em\raise-1.1ex\hbox{$\sim$}}\;}}

\newcommand{\lsim}
{{\;\raise0.3ex\hbox{$<$\kern-0.75em\raise-1.1ex\hbox{$\sim$}}\;}}

\def\lesq{\ell_e^2}
\def\resq{r_e^2}

\def\lfsq{\ell_f^2}
\def\rfsq{r_f^2}

\begin{document}

\begin{flushright}
{IISc-CHEP/03/09}
\end{flushright}
\begin{center}
\textbf{\Large Use of transverse beam polarization to probe anomalous 
\\\vspace*{0.3cm}
$VVH$ interactions at a Linear Collider} 
\end{center}
\vspace*{0.1cm}
\begin{center}
{\mbox{\large Sudhansu~S.~Biswal and Rohini~M.~Godbole}}\\ \vspace*{0.2cm}
{\small \it Centre for High Energy Physics, Indian Institute of Science,
Bangalore 560 012, India} 
\end{center}
\vspace*{0.1cm}

\begin{abstract}
We investigate use of transverse beam polarization 
in probing anomalous coupling of a Higgs boson to a pair of 
vector bosons, at the International Linear Collider (ILC). 
We consider the most general form of $VVH$ ($V = W/Z$) vertex 
consistent with Lorentz invariance and investigate its effects 
on the process $e^+ e^- \rightarrow f \bar f H$, $f$ being a 
light fermion. Constructing observables with definite $CP$ and 
naive time reversal ($\tilde T$) transformation properties, 
we find that transverse beam polarization  helps us to improve 
on the sensitivity of one part of the anomalous $ZZH$ coupling that is 
odd under $CP$. Even more importantly it provides the 
possibility of discriminating from each other, two terms 
in the general $ZZH$ vertex, both of which are 
even under $CP$ and $\tilde T$. Use of transverse beam polarization
when combined with information from unpolarized and linearly polarized
beams therefore, allows one to have completely independent probes
of all the different parts of a general $ZZH$ vertex. 
\end{abstract}

\section{\label{sec:intro}Introduction}
The Standard Model (SM) has been incredibly successful in
describing all the available experimental data. However, the
Higgs mechanism responsible for generating masses of all
the particles in the SM still lacks direct testing. The SM 
predicts  existence of one Higgs boson, a spin-0 particle,
even under charge conjugation and parity ($C, P$)
transformation, whereas scenarios beyond the SM usually 
imply more than one Higgs
boson with different $CP$ and weak isospin quantum 
numbers~\cite{HHG,abdelrev}. 
For example, the Minimal Supersymmetric extension of the Standard Model
(MSSM)~\cite{MSSMbook} consists of five
Higgs particles: two $CP$-even neutrals, a $CP$-odd neutral and
a pair of charged scalars~\cite{HHG,abdelrev,MSSMbook}. 
Therefore, search for the Higgs boson
and study of its various properties is one of the major goals of all
the current and future colliders~\cite{Godbole:2002mt}.

Direct searches at the LEP collider have put a lower bound on the mass of 
the SM Higgs boson: $m_H > 114.4$ GeV~\cite{higmin}, whereas the LEP
electroweak precision measurements have been used to put an upper bound on its
mass, of  about 185 GeV~\cite{higmax} at 95\% confidence level (CL). 
Theoretical constraints on the mass of the SM 
Higgs boson exist, given by demanding unitarity  of scattering 
amplitudes, perturbativity of Higgs self-coupling, stability 
of the electroweak vacuum and no fine-tuning in the radiative 
corrections in the Higgs sector~\cite{abdelrev,Godbole:2002mt}.
The Large Hadron Collider (LHC) is expected~\cite{LHC} to be capable of
searching for the SM Higgs boson over the mass range allowed by all the above 
considerations, viz. 114.4 GeV $\lsim  m_H \lsim $ 800 GeV. 
After the discovery of the Higgs boson, precise determination 
of its interactions with other particles  would be necessary to establish 
it as {\it the} SM Higgs boson. A detailed study of Higgs sector may also 
provide hints for new physics beyond the SM. Hence high precision
measurements at the International $e^+e^-$ Linear Collider
(ILC)~\cite{ILC,RDR} and its combination with the information available
from the LHC, will be required to establish the nature of Higgs 
boson~\cite{Weiglein:2004hn} in general and its CP property in 
particular~\cite{Godbole:2004xe}. 
Part of this  exercise could be achieved by determining the tensor 
structure of the coupling of the spin-0 state with different SM particles. 
The determination of the CP nature of the Higgs boson at various colliders has 
been a subject of many investigations~\cite{cpnsh}. 
Various kinematical distributions for the process $e^+ e^- \to f \bar f H$,
proceeding via vector boson fusion and Higgs-strahlung, can be used
to probe the $ZZH$ vertex~\cite{dist,Hagiwara:1993sw}.
At an $e^+e^-$ collider, the angular  and energy distribution of the $Z$
boson in the process $e^+ e^- \rightarrow Z H$ can provide information
about the $ZZH$ coupling~\cite{Barger:1993wt, CP-full}. Further 
the tensor structure
of the $ZZH$ and $t \bar t H$ coupling may be probed in a model independent
way by looking at the shapes of the threshold excitation curve in the
processes $e^+e^- \rightarrow Z H$~\cite{Miller:2001bi} and
$e^+ e^- \rightarrow t \bar t H$~\cite{bhupal} respectively. 
Higher order contributions in a renormalizable theory~\cite{kniehl:NPB}
or higher dimensional operators in an effective 
theory~\cite{Buchmuller:1985jz,operator}
may give rise to anomalous parts of $VVH$ vertex that we 
discuss here. In the context
of higher dimensional operators the anomalous $VVH$ vertex
has been studied in
Refs.~\cite{ Hagiwara:1993sw,Rattazzi:1988ye,He:2002qi,Hagiwara:1993ck,
Gounaris:1995mx,Hagiwara:2000tk,Chang:1993jy,Han:2000mi,Biswal:2005fh,
Biswal:2008tg,Rindani:2009pb,Grzadkowski:1995hi,Kilian:1996wu}. 
Particularly in Ref.~\cite{Hagiwara:2000tk} the authors have made
a detailed analysis using the optimal observable
technique~\cite{Atwood:1991ka}, to probe $ZZH$ and $\gamma ZH$ couplings,
whereas Refs.~\cite{Chang:1993jy,Han:2000mi,Biswal:2005fh,Biswal:2008tg}
use asymmetries constructed using differences in the kinematical
distributions of the decay products.
Similar methods can be used to probe the $ZZH$ vertex at the LHC as 
well~\cite{cpnsh,Godbole:2007cn}.  
The anomalous $VVH$ couplings can also be 
studied at an $e \, \gamma$ collider~\cite{egamma}. Previous 
studies~\cite{Hagiwara:2000tk,Han:2000mi,Biswal:2005fh,Biswal:2008tg} 
showed that the use of linear beam polarization and as well as measurement
of the polarisation of the final state fermion for the case of the 
$\tau$, improves the sensitivity 
of these asymmetries to some of these  anomalous couplings 
substantially and provides independent probes of many of the parameters 
characterizing them. However, a few ambiguities still remain.  

%
It has been realized that at future $e^+e^-$ linear colliders 
spin rotators can be used to obtain transverse beam polarization.
In the context of the ILC, use of transverse polarization
in probing new physics, including anomalous $VVH$ vertex,
has been discussed in 
Refs.~\cite{Rindani:2009pb,MoortgatPick:2005cw,
ref_trans1,ref_trans2,ref_trans3,Rao:2006hn,ref_trans4}. 
In this note we study how  information obtained using transverse
polarization  can improve the situation and make possible completely
independent determination of the different parts of the $ZZH$ anomalous 
coupling.  
The asymmetries with unpolarized beams suffer from a suppression factor
($\lfsq - \rfsq$), where
$l_f$ ($r_f$) is the left-(right-)handed coupling
of the fermion to the $Z$ boson~\cite{Biswal:2005fh}.  
With polarized beams using the spin direction of the beam,  
construction of asymmetries 
which may not suffer from this suppression factor, becomes possible. 
In this note we explore use of transverse beam polarization 
to probe $VVH$ vertex.

In the next section we discuss possible sources of 
anomalous $VVH$ couplings  and construct  observables, specific to the 
use of transverse polarization.  In sec.~\ref{sec:results} we present 
numerical analysis of the same.
We summarize our results in Sec.~\ref{sec:conclude}.
%
%

\section{\label{sec:obs} General structure and Probes of 
$VVH$ couplings }
\begin{figure}[!h]
\begin{center}
\epsfig{file=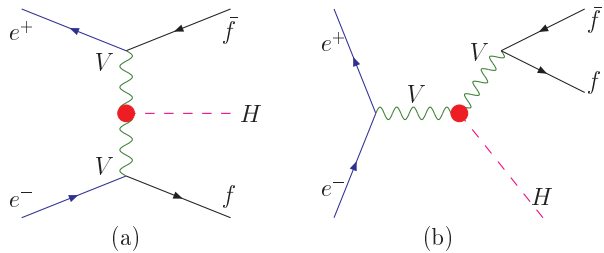,width=8.5cm}
\caption{\label{fig:higgs-prod}Feynman diagrams
for the process $e^+e^- \to f\bar f
H$; (a) is t-channel or fusion diagram, while (b) is s-channel
or Bjorken diagram. For $f=e,\nu_e$ both (a) and (b) contribute
whereas for all the other fermions only (b) contributes.}
\end{center}
\end{figure}

At an $e^+e^-$ collider the dominant Higgs production process,
{\it viz} $e^+e^- \to f \bar f H $ where $f$ is any light fermion,
proceeds via the $VVH$ interactions with $V = W, Z$. Processes with
electrons and $\nu_e$'s in the final states receive contribution 
from both the s- and t-channel diagrams (Figs.~\ref{fig:higgs-prod}).

The interaction term involving the Higgs boson and a pair of
gauge bosons arises from the Higgs kinetic term in the SM/MSSM
Lagrangian.
However, once we accept the SM
to be only an effective low-energy theory, higher dimensional (and
hence non-renormalizable) terms are allowed.
The most general $VVH$ vertex, consistent with Lorentz invariance,
can be written as
\begin{eqnarray}
\Gamma_{\mu\nu} &=& g_V\left[a_V \ g_{\mu\nu}+\frac{b_V}{m_V^2}
(k_{1\nu} k_{2\mu}
 - g_{\mu\nu}  \ k_{1} \cdot k_{2})
+\frac{\tilde b_V}{m_V^2} \ \epsilon_{\mu\nu\alpha\beta} \
k_1^{\alpha}  k_2^{\beta}\right], 
\label{eq:coup}
\end{eqnarray}
where $k_i$ denote the momenta of the two $W$'s ($Z$'s). Here
\[
g_W^{SM} = e \, \cot\theta_W M_Z  \ , \qquad
g_Z^{SM}= 2 \, e M_Z/\sin2\theta_W,
\]
$\theta_W$ being the weak-mixing angle and
$\epsilon_{\mu\nu\alpha\beta}$ the antisymmetric tensor with
$\epsilon_{0123}~=~1$.
In the SM, at tree level $a_Z=a_W=1$ and $b_V=\tilde b_V=0$.

In general, each of the couplings of Eq.~\ref{eq:coup} 
($a_V,~b_V$ and $\tilde b_V$) can be complex. 
However, for all the observables that we construct 
for the process $e^+ e^-
\to f \bar f H$, one overall phase can always be rotated away. We
choose $a_Z$ to be real and allow the others to be complex. 
Further, we assume $a_Z$ and $a_W$ to be close to their SM value
i.e. $a_V = 1 + \Delta a_V$.
The most general coupling of a $CP$-even Higgs boson with
a pair of vector bosons is expressed by the terms containing
$a_V$ and $b_V$ in Eq.~\ref{eq:coup} whereas the $CP$-odd one
corresponds to $\tilde b_V$. 
Simultaneous presence of both sets of terms would indicate
$CP$-violation. Note that a non-vanishing value for either $\Im(b_V)$ or
$\Im(\tilde b_V)$ destroys the hermiticity of the effective theory.

In an effective theory which satisfies
${\rm SU(2)_L \otimes U(1)_Y}$ symmetry, couplings $b_V$ and
$\tilde b_V$ can be realized as lowest order corrections arising from
dimension-six operators such as $F_{\mu \nu} F^{\mu \nu} \Phi^\dagger \Phi$
or $F_{\mu \nu}\tilde{F}^{\mu \nu} \Phi^\dagger \Phi$ where $\Phi$
is the usual Higgs doublet, $F_{\mu \nu}$ the field strength tensor
and $\tilde{F}_{\mu \nu}$ its dual~\cite{operator}. 
The coupling constants
may even have non-trivial momentum dependence (form-factor behavior).
However, for a theory with a cut-off scale $\Lambda$ large compared
to the energy scale at which the scattering experiment is to
be performed, momentum dependence  would be very weak and
can be neglected for our study. Hence we treat
$\Delta a_V$, $b_V$ and $\tilde b_V$ as phenomenological,
energy-independent parameters and keep terms up to linear order in 
our analysis, assuming them to be small 
compared to the SM coupling.
\begin{table}[!h]
\begin{center}
\begin{tabular}{c|ccccc|}
\cline{2-6}
\\[-2.5ex]
\multicolumn{1}{c|}{} & $a_V$  & $\Re(b_V)$ & $\Im(b_V)$ &
$\Re(\tilde b_V)$ & $\Im(\tilde b_V)$\\
\hline
\multicolumn{1}{|c|}{$CP$} & $+$  & $+$ & $+$& $-$ & $-$ \\
\multicolumn{1}{|c|}{$\tilde T$} & $+$  & $+$ & $-$& $-$ & $+$
\\
\hline
\end{tabular}
\end{center}
\vskip -0.4cm
\caption{\em Transformation properties of
the various operators (identified by their coefficients) in the
effective Lagrangian.}
\vspace{-0.3cm}
\label{tab:coup}
\end{table}
Table~\ref{tab:coup} shows the $CP$ and $\tilde T$ transformation
properties of various operators in the effective Lagrangian corresponding
to different anomalous couplings given in the table.
In principle, the full process $e^+ e^- \rightarrow H f \bar f$ can   
receive contributions from some additional higher dimensional 
operators~\cite{Buchmuller:1985jz}, which could arise (say)   
from a $Z'$ exchange in an effective 
theory~\cite{Grzadkowski:1995hi,Kilian:1996wu}. Under the assumption 
of flavour universality these additional contributions to the 
$e^+ e^- \rightarrow f \bar f H$ amplitude, will have the same structure 
as that coming from some of the terms in our anomalous $VVH$ vertex. Thus 
our present analysis includes such contributions as well, subject to  
the above flavour universality assumption.

As was the case with our earlier analyses~\cite{Biswal:2005fh,Biswal:2008tg} 
we construct various kinematical quantities ${\cal C}_i$ as
combinations of the available different particle momenta and
their spins.
Then we define observables $O^{\rm T}_{i}$, as expectation values
of the signs of ${\cal C}_i$, i.e.  $O^{\rm T}_{i} = \langle {\rm
sign}({\cal C}_i)\rangle$, each of which
have well-defined $C$, $P$ and $\tilde T$ transformation properties.
More explicitly $O^{\rm T}_{i}$'s can be expressed as: 
\bea
\label{obs_expt}
O^{\rm T}_{i} &=& \frac{1}{\sigma_{\rm SM}}\, 
\int [{\rm sign}({\cal C}_{i})] 
\frac{d \sigma}{ d^3 p_{_H}d^3p_{_f}}\, d^3 p_{_H}d^3p_{_f} \\   
&=& \frac {\sigma ({\cal C}_i > 0) - \sigma ({\cal C}_i < 0)} 
{\sigma_{\rm SM}}.\nonumber
\eea
Within the aforementioned linear approximation,  $O^{\rm T}_{i}$'s 
may be used to probe the contribution of specific operator(s) in the 
effective Lagrangian with the same $CP$ and $\tilde T$ transformation
properties. Of course, by construction this seems to make it  
impossible to have completely independent 
probes for those  parts of the anomalous $VVH$ vertex which have 
the  same $CP$ and $\tilde T$ properties : viz.   $\Re(b_V)$ and $a_V$.  
Such was the case with the observables we constructed earlier 
with unpolarized and longitudinally polarized 
beams~\cite{Biswal:2005fh,Biswal:2008tg}. However, 
we find  that use of transverse beam polarization 
allows us to distinguish the contributions from 
$\Re(b_Z)$ and $a_Z$ from each other. As a result, using information 
from unpolarized and transversely/linearly polarized 
beams, one can determine, completely independently, all 
the parts of the general $ZZH$ coupling. 
Similar observation has also been made recently 
in the context of general 
$ZZH$ and $Z\gamma H$ three point interactions in 
Ref.~\cite{Rindani:2009pb}. 
\begin{table}[!h]
\begin{center}
\begin{tabular}{|r|l|ccccc|c|c|c|}
\hline
\hline
&&&&&&&&\\[-3mm]
ID&${\cal C}_i$ & {$C$} & {$P$} & {$CP$} & {$\tilde T$} &
{$CP \tilde T$} &
$\begin{array}{c}
\mbox{Observable}~(O^{\rm T}_{i})
\end{array}$
& \mbox{Coupling} \\
[2mm]
\hline
\hline
&&&&&&&&\\[-3mm]
1& $ {({\vec{p}}{_{_H}})}_x^2 \ - \ {({\vec{p}}{_{_H}})}_y^2 $
& {$+$} & {$+$} & {$+$} & {$+$} & {$+$} &$O^{\rm T}_1$& $a_V,~\Re(b_V)$ \\
[2mm]
\hline
\hline
&&&&&&&&\\[-3mm]
2& $ ({\vec{P}}{_f})_x * ({\vec{P}{_f})_y} * ({\vec{p}}{_{_H}})_z$
& {$+$} & {$-$} & {$-$} & {$-$} & {$+$} &$O^{\rm T}_2$& 
$\Re(\tilde b_Z)$ \\
[2mm]
\hline
&&&&&&&&\\[-3mm]
3& $ ({\vec{p}}{_{_H}})_x * ({\vec{p}}{_{_H}})_y * ({\vec{P}}{_f})_z$
& {$-$} & {$-$} & {$+$} & {$-$} & {$-$} &$O^{\rm T}_3$& 
$\Im(b_Z)$ \\
[2mm]
\hline
\end{tabular}
\caption{ \label{tab:finalcorr}
{\em Various possible ${\cal C}_i$'s, their
transformation properties, the associated observables $O^{\rm T}_{i}$
and the anomalous couplings on which they provide information.
Here, $\vec{P}_f \equiv \vec{p}_{_f} - \vec{p}_{_{\bar f}}$ 
 and ${\vec{p}}{_{_H}}$ is the momentum of Higgs boson (to be deduced from
the measurement of its decay products).}}
\end{center}
\end{table}

Table~\ref{tab:finalcorr} lists the observables 
$O^{\rm T}_{i}$ ($i$ = $1$--$3$), specific to the case of transverse 
polarization,  along with 
their transformation properties and the anomalous coupling they may 
constrain.  A brief description of  the same follows:

\begin{description}
\item
[1. $O^{\rm T}_1$] is an observable defined as 
the difference in partially integrated cross
sections normalized by the SM cross section and is given by 
\begin{eqnarray}
O^{\rm T}_1 
&=& \frac {\sigma ({\cal C}_1 > 0) - \sigma ({\cal C}_1 < 0)}
{\sigma_{\rm SM}} \\ \nonumber
&=& \frac{\sigma (\cos2\phi_H>0) - \sigma (\cos2\phi_H<0)}
{\sigma_{\rm SM}},
\label{obs_O5_expt}
\end{eqnarray}
where $\phi_H$ is the azimuthal angle of the Higgs boson  
with respect to the XZ-plane defined by taking  
Z-axis to be direction of initial state $e^-$ 
momentum and choosing positive X-axis to be the direction of its 
transverse polarization. 
This observable is derived from  
\begin{eqnarray}
{\cal C}_1 &\equiv& [{({\vec{p}}{_{_H}})}_x^2 \ - \ {({\vec{p}}{_{_H}})}_y^2]  
\ (\propto \cos2\phi_H). 
\label{eq:c1}
\end{eqnarray}
One can see that this is even under $CP$, $\tilde T$ and 
hence can be sensitive to $a_V$ 
and $\Re(b_V)$. 
\item
[2. $O^{\rm T}_2$] can be constructed 
using ${\cal C}_2$, here
\bea
{\cal C}_2 &=& ({\vec{P}}{_f})_x * ({\vec{P}{_f})_y} *
({\vec{p}}{_{_H}})_z \\ \nn
           &\propto& 
[\vec{S_e}\cdot \vec{P_f}]*
                     [(\vec{S_e} \times \vec{P_e})\cdot \vec{P_f}]*
                     [\vec{P_e}\cdot \vec{p_H}],
\label{eq:c2}
\eea
and 
$\vec{P}_e \equiv \vec{p}_{e^-} - \vec{p}_{e^+}$,
$\vec{P}_f \equiv \vec{p}_{_f} - \vec{p}_{_{\bar f}}$,
$\vec{S}_e \equiv \vec{s}_{e^-} - \vec{s}_{e^+}$.

Hence $O^{\rm T}_2$ is $CP$-odd, $\tilde T$-odd and thus would be 
proportional to $\Re(\tilde b_Z)$. It can be expressed as 
\bea
O^{\rm T}_2 &=& 
\frac {\sigma ({\cal C}_2 > 0) - \sigma ({\cal C}_2 < 0)}
{\sigma_{\rm SM}}.
\label{obs_O6_expt}
\eea
\item
[3. $O^{\rm T}_3$] is derived from ${\cal C}_3$,
where 
\bea
\label{eq:c3}
{\cal C}_3 &=& ({\vec{p}}{_{_H}})_x * ({\vec{p}}{_{_H}})_y * 
({\vec{P}}{_f})_z  \\ \nn
           &\propto& 
[\vec{S_e}\cdot {\vec{p}_H}]*
                     [(\vec{S_e} \times \vec{P_e}) \cdot \vec{p}_H]*
                     [\vec{P_e}\cdot \vec{P_f}]
\eea
is $CP$-even and $\tilde T$-odd. $O^{\rm T}_3$ is defined as 
\bea
O^{\rm T}_3 &=& 
\frac {\sigma ({\cal C}_3 > 0) - \sigma ({\cal C}_3 < 0)}
{\sigma_{\rm SM}}
\label{obs_O7_expt}
\eea
and may be used to probe $\Im(b_Z)$. 
\end{description}
It is obvious from the structure of ${\cal C}_{1}-{\cal C}_{3}$ 
that $O^{\rm T}_1$, $O^{\rm T}_2$ and $O^{\rm T}_3$ 
will be nonzero only for transversely polarized beams. Out 
of these three observables the last two 
require charge measurement of the final state particles and
hence cannot be used for final states with light 
quarks and neutrinos.

%
%

\section{\label{sec:results}Numerical results}

In the numerical analysis, we focus on the intermediate mass Higgs boson
($2m_b \leq m_H \leq 140$GeV), for which $H\to b\bar b$ is the
dominant decay mode with a branching fraction $\simeq 0.68$.
We consider the case of the ILC operating at
a center of mass energy of 500 GeV, with a  Higgs boson mass
120 GeV.  Further,  we take the $b$-tagging efficiency to be 0.7.
We impose various kinematical cuts to suppress dominant backgrounds to
the process under study, viz. $e^+e^- \to f \bar f H \to f \bar f b\bar b$. 
To begin with, we demand that each of the final state particles
has energy above a minimum energy, as well as a minimum angular deviation
from the beam pipe. Furthermore, they should be well separated from each 
other.  For final state with neutrinos, we need  to impose a cut on missing 
transverse momentum. The specific kinematic cuts we impose are:
\begin{eqnarray}
\begin{array}{rclcl}
E_f &\ge&  10 \,\mbox{GeV} &&
\mbox{for each visible outgoing fermion}\\[0.2cm]
5^\circ~ \le ~ \theta_0 &\le& 175^\circ& &
\mbox{for each visible outgoing fermion}\\[0.2cm]
p_T^{\mbox {miss}} & \geq & 15\, \mbox{GeV} & &
\mbox{for events with} ~\nu'\mbox{s} \\[0.2cm]
\Delta R_{j j} & \geq & 0.7 & &
\mbox{for each pair of jets}\\[0.2cm]
\Delta R_{\ell\ell} & \geq & 0.2 & &
\mbox{for each pair of charged leptons}\\[0.2cm]
\Delta R_{l j}&\ge& 0.4&&
\mbox{for jet-lepton isolation },
\end{array}
\label{eq:cuts}
\end{eqnarray}
where $(\Delta R)^2 \equiv (\Delta \phi)^2 + (\Delta \eta)^2 $,
$\Delta \phi $ and $\Delta \eta$ being the separation between the two
entities in azimuthal angle and rapidity respectively.

Additional cuts may be imposed on the
invariant mass of the $f\bar f$ system to enhance(suppress) the
contributions coming from s-channel $Z$-exchange process, namely
\begin{equation}
\begin{array}{rcl}
R1 &\equiv& \left| m_{f\bar f} - M_Z \right| \leq 5 \, \Gamma_Z \hspace{0.5cm}
\mbox{ $\Longrightarrow$ select \ $Z$-pole} \ ,\\[1ex]
R2 &\equiv& \left| m_{f\bar f} - M_Z \right| \geq 5 \, \Gamma_Z \hspace{0.5cm}
\mbox{ $\Longrightarrow$ de-select \ $Z$-pole} \ ,
\end{array}
 \label{cuts:Z}
\end{equation}
where $\Gamma_Z$ is the width of the $Z$ boson.
Alternatively, for $\nu\bar\nu H$ final state, the same goal can
be achieved by demanding
\begin{equation}
\begin{array}{rcl}
 R1^\prime &\equiv & E^-_H \leq E_H \leq E^+_H,  \nonumber
\\[1ex]
R2^\prime &\equiv & E_H < E^-_H \ \mbox{or} \ E_H > E^+_H.
\end{array}
\label{cuts:EH}
\end{equation}
Here $E_H^\pm = (s + m_H^2 - (m_Z\mp5\Gamma_Z)^2) /
(2\sqrt{s})$.

We construct various $O^{\rm T}_i, i=1,3$ as the  expectation values of 
the sign of ${\cal C}_{i}, i=1,3$ as mentioned earlier. 
It was pointed out in Refs.~\cite{Hikasa:1984ng,Hikasa:1985qi} that in 
the massless limit of electron, transverse beam polarization does not 
affect the azimuthally integrated SM cross-section, due to the electronic
chiral symmetry of the SM.
The anomalous $VVH$ vertex that we are considering does not 
change the chiral structure of the theory. Hence 
transverse beam polarization also cannot alter 
contributions to the total rate, 
coming from the anomalous parts of the $VVH$ vertex;  
viz. $\Re(b_Z),~\Delta a_Z$.  
Further to our choice of $XZ$ plane, 
we assume direction of $e^+$ 
polarization to be opposite to that of $e^-$ and take  $e^-$, $e^+$ 
polarization (${{\cal P}_{e^-}^T}$, ${{\cal P}_{e^+}^T}$) to be 
80\% and 60\% respectively. We denote  $\theta_H$, $\phi_H$ as  
the polar and azimuthal angle of the 
Higgs boson and $\phi_f$  as the angle between  
production planes of the Higgs boson and final state fermion.  

The statistical fluctuation in the cross section and that in an asymmetry,
for a given luminosity ${\cal L}$ and fractional systematic error $\epsilon$ ,
can be written as:
\begin{equation}
\Delta\sigma = \sqrt{\sigma_{SM}/{\cal L} + \epsilon^2\sigma_{SM}^2} \  \ ,
\label{Dsig}
\end{equation}
\begin{equation}
(\Delta A)^2= \frac{1-A^2_{SM}}{\sigma_{SM}{\cal L}} + \frac{\epsilon^2}{2}(1-
A^2_{SM})^2, 
\label{Dasym}
\end{equation}
where $\sigma_{SM}$ and $A_{SM}$ are the SM value of cross section and
asymmetry respectively.
Since all the anomalous couplings are kept only up to linear order, 
any of the observables; total rate or asymmetries, can be written as:
\[ {\cal O}(\{{\cal B}_i\}) = \sum \ O_i \ {\cal B}_i, \]
where ${\cal B}_i$ is the anomalous coupling.  
We calculate the possible limits of sensitivity 
on the  anomalous  couplings by demanding
that $|{\cal O}(\{{\cal B}_i\})-{\cal O}(\{0\})|\le f \ \Delta{\cal O}$.
Here  $f$ is the degree of statistical significance,
${\cal O}(\{0\})$ is the SM value of ${\cal O}$ and $\Delta{\cal O}$ 
is the statistical fluctuation in ${\cal O}$. In our analysis we 
consider $\epsilon$ = 0.01, $\cal L$ = 500 fb$^{-1}$ and $f$ = 3.
Next we  present our numerical results  
for the case of $ZZH$ and $WWH$ couplings respectively.
\subsection{\label{sec::zzh}Anomalous $ZZH$ Couplings}
We observe that transverse beam polarization will lead to 
additional contributions to the differential cross section only 
when both the beams are polarized. 
The additional contribution from transverse beam
polarization in the squared matrix element (MESQ) is 
proportional to the interference 
between different helicity amplitudes  and hence these additional terms
are proportional to ${{\cal P}_{e^-}^T} {{\cal P}_{e^+}^T}$, 
where ${{\cal P}_{e^\mp}^T}$ is polarization of $e^\mp$. 
This proportionality factor
(${{\cal P}_{e^-}^T} {{\cal P}_{e^+}^T}$)
can be understood as a consequence of electronic chiral symmetry mentioned
earlier ~\cite{Hikasa:1985qi}.

Use of transverse beam polarization to probe anomalous 
$ZZH$ and $Z \gamma H$ couplings had been addressed in 
Refs.~\cite{Hagiwara:1993sw,Rindani:2009pb}.
In Ref.~\cite{Hagiwara:1993sw} it was pointed out that 
polarization does not provide qualitatively new 
information about the anomalous $ZZH$ couplings if they 
are assumed to be real. This is no longer true once the 
couplings are allowed to be complex. 
A recent discussion~\cite{Rindani:2009pb} 
shows how the anomalous $ZZH$ and $Z \gamma H$ vertex may be 
probed with  transversely polarized beams using  $Z$-angular distribution.   
However, not all the parts of the anomalous couplings are accessible
in this process. For example, $\Im(b_Z)$ and $\Re(\tilde b_Z)$ 
cannot be probed using the  $Z$-angular distribution. 
Our analysis goes beyond that in  both of the 
Refs.~\cite{Hagiwara:1993sw,Rindani:2009pb},  in that we 
consider the full process $e^+e^- \to f \bar f H$ and with the 
most general form of anomalous $ZZH$ vertex, allowing the couplings to
be complex. 
  
The structure of ${\cal C}_{1}$, ${\cal C}_{2}$ and ${\cal C}_{3}$  given by
Eqs. \ref{eq:c1}--\ref{eq:c3}  
indicates that the corresponding observables,
$O^{\rm T}_{1}$, $O^{\rm T}_{2}$ and $O^{\rm T}_{3}$ 
will be nonzero only for transversely polarized beams.
This  can be understood by noticing that  
the transverse polarization of $e^-$ (along 
positive X-axis) and beam direction (along Z-axis) define a preferred, fixed 
XZ-plane, whereas for unpolarized or longitudinally polarized beams  
such a plane does not exist. 
Thus these  three observables are unique for transversely polarized beams.  
Final states containing light quarks and muons receive contribution 
only from the $s$-channel Higgs-strahlung process,  
whereas those with electrons receive contribution from 
the $t$-channel diagram as well. $O^{\rm T}_{2}$ and $O^{\rm T}_{3}$  
require charge measurement of final state particles. Hence we do not use  
final states with quarks while assessing their efficacy 
to probe the anomalous couplings. 

For transverse beam polarization the contributions coming 
from different anomalous parts to MESQ have different azimuthal angular 
dependencies.  
Since ${\cal C}_{1} \propto   \cos2\phi_H $ and  is even under both  
$CP$- and $\tilde T$-transformation; the corresponding observable    
$O^{\rm T}_{1}$ can then in principle involve terms proportional to  
both $\Delta a_Z$ and $\Re (b_Z)$.
\begin{figure}[!h]
\begin{center}
\epsfig{file=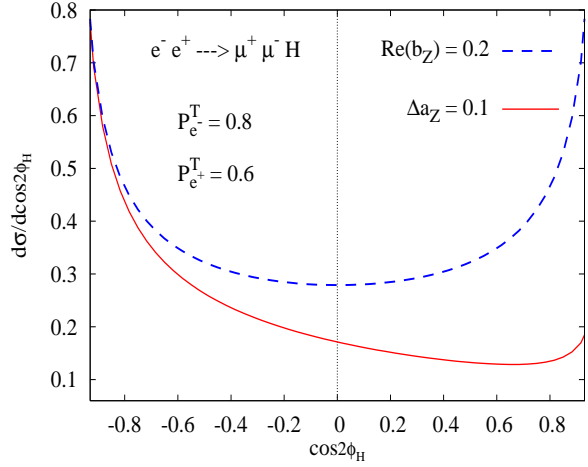,width=8.5cm,height=6.5cm}
\caption{\label{fig:az-rebz-tr}{\em Azimuthal angle distribution 
of the Higgs boson for 80\%, 60\% transversely polarized $e^-,  e^+$ beams .}}
\end{center}
\end{figure}
Fig.~\ref{fig:az-rebz-tr} displays differential cross section 
($d {\sigma} / d \cos2\phi_H $) as a function of $\cos2\phi_H$ 
for transversely polarized $e^-/e^+$ beams   
and nonzero values of these two $CP$- and 
$\tilde T$-even anomalous $ZZH$ couplings.  
We note that the observable $O^{\rm T}_{1}$ does not receive 
any nonzero contribution from $\Re (b_Z)$ and thus is an 
independent probe of $\Delta a_Z$ alone. 
We will get back to this issue after discussing the possible sensitivity
of $O_1^{\rm T}$ as a probe of $\Delta a_Z$. 

Imposing various kinematical cuts of Eqs.~\ref{eq:cuts}, along-with
the $R1$-cut of Eq.~\ref{cuts:Z} to select $Z$-pole contribution, 
$O^{\rm T}_{1}$ for muon and quark final states, with transverse
polarization (${{\cal P}_{e^-}^T}$, ${{\cal P}_{e^+}^T}$) of the initial beams,
can be written, keeping terms up to linear order in anomalous couplings, as 
\begin{equation}
\begin{array}{rcl}
O^{\rm T}_{1} (R1-\rm{cut}) &=&
\left\{
\begin{array}{lcl}
\displaystyle
\frac{[{{\cal P}_{e^-}^T} {{\cal P}_{e^+}^T}]~
[- 0.37\ (\ 1 \ + \ 2 \ \Delta a_Z)]}{0.86}
&  & (\mu^+ \mu^- H)\\[3.0ex]
\displaystyle
\frac{[{{\cal P}_{e^-}^T} {{\cal P}_{e^+}^T}]~
[- 5.7\ (\ 1 \ + \ 2 \ \Delta a_Z)]}{13.2}
&  & (q \bar q H)\\[3.0ex]
\end{array}
\right.
\end{array}
\label{eq:asym_O3T}
\end{equation}
Here, each of the numerical factors are in
femtobarns with the denominator being the SM cross
section. It is clear from the above expression that $O^{\rm T}_{1}$ is 
nonzero only when both the beams are transversely polarized, as 
it should be. 

Taking $e^-$ and $e^+$ beam polarization to be 80\% and 60\% respectively,
using the final state with $\mu$ and quarks, lack of deviation from the SM 
value of 
$O^{\rm T}_{1}$, at $3 \sigma$ level,  would limit $\Delta a_Z$ to:
\bea
|\Delta a_Z| & \lsim & 0.1 
\quad {\rm for} \,\, {\cal L} = 500\, {\rm fb}^{-1}.
\label{lim-deaz-m}
\eea

The best possible sensitivity limit on $\Delta a_Z$ obtained for 
the unpolarized case in our earlier analysis~\cite{Biswal:2005fh} 
was: $|\Delta a_Z| \leq  0.04$. 
However, the observable (total cross section with unpolarised beams) used 
to obtain this bound on $\Delta a_Z$ also 
gets contribution from $\Re(b_Z)$~\cite{Biswal:2005fh}.  

We emphasize here that though $O^{\rm T}_{1}$ is 
even under $CP$- and $\tilde T$-transformation,   
it receives contribution only from $\Delta a_Z$  
(and not from $\Re(b_Z)$), as shown by azimuthal angular distribution 
displayed in Fig~\ref{fig:az-rebz-tr}.  
Construction of such an observable was not possible with 
unpolarized beams~\cite{Biswal:2005fh}. In fact a similar  observation 
has been made recently in Ref.~\cite{Rindani:2009pb} in the context of  
the process $e^+ e^- \to Z H$. The additional contribution to the MESQ
for this  process due to the transverse beam polarization proportional 
to $\Re(b_Z)$, vanishes identically as  these terms are proportional to 
$\vec{s}_{e^-} . \vec{p}_{e^-}$, $\vec{s}_{e^+} . \vec{p}_{e^+}$.
For the full process $e^+ e^- \to f \bar f H$,  
even  though the additional terms  in the MESQ due to transverse polarisation  
are nonzero, the contribution to $O^{\rm T}_{1}$ proportional
to $\Re (b_Z)$ constructed after integrating over the remaining phase space 
variables vanishes.  

Furthermore, using ${\cal C}_{1}$ we can construct an azimuthal asymmetry 
defined as
\begin{eqnarray}
{\cal A}_{1}^{\rm T} &=& 
\frac{\sigma (\cos2\phi_H>0) - \sigma (\cos2\phi_H<0)}
{\sigma (\cos2\phi_H>0) + \sigma (\cos2\phi_H<0)}\ 
\label{obs_Ot1_assym}
\end{eqnarray}
With $R1$-cut expression for this asymmetry
for muon and quark final states is given by
\begin{equation}
\begin{array}{rcl}
{\cal A}^{\rm T}_{1} (R1-\rm{cut}) &=&
\left\{
\begin{array}{lcl}
\displaystyle
\frac{[{{\cal P}_{e^-}^T} {{\cal P}_{e^+}^T}]~
[- 0.37\ (\ 1 \ + \ 2 \ \Delta a_Z)]}
{[0.86\ (\ 1 \ + \ 2 \ \Delta a_Z) \ + \ 8.2 \ \Re(b_Z)]}
&  & (\mu^+ \mu^- H)\\[3.0ex]
\displaystyle
\frac{[{{\cal P}_{e^-}^T} {{\cal P}_{e^+}^T}]~
[- 0.57\ (\ 1 \ + \ 2 \ \Delta a_Z)]}
{[1.32\ (\ 1 \ + \ 2 \ \Delta a_Z) \ + \ 12.5 \ \Re(b_Z)]}
&  & (q \bar q H)\\[3.0ex]
\end{array}
\right. 
\\ \\
&\simeq&
-~0.43~\left[{{\cal P}_{e^-}^T} {{\cal P}_{e^+}^T}\right]~
\left[ 1 - 9.5 \ \Re(b_Z)\right]
\end{array}
\label{eq:asym-O1t}
\end{equation}
The second expression of Eq.~\ref{eq:asym-O1t} is written to linear order
in anomalous couplings. Since the numerator of Eq.~\ref{obs_Ot1_assym}
does not have any contribution linear in $\Re(b_Z)$, the asymmetry of
Eq.~\ref{eq:asym-O1t} involves only $\Re(b_Z)$ to linear order. 
${\cal A}_{1}^{\rm T}$ with final states $\mu$'s and light quarks, for an
integrated luminosity ${\cal L} = 500\, {\rm fb}^{-1}$, leads to
a $3 \sigma$ constraint of
\bea
|\Re(b_Z)| & \lsim & 0.021.
\eea
Thus we see that 
both the $CP$- and $\tilde T$-even
couplings, $\Re(b_Z)$ and $\Delta a_Z$, can be probed
independently using ${\cal A}_{1}^{\rm T}$ and $O^{\rm T}_{1}$
respectively, which was not possible earlier with unpolarized
and/or linearly polarized beams~\cite{Biswal:2005fh,Biswal:2008tg}. 
It is interesting to note that 
this sensitivity limit on $\Re(b_Z)$ obtained using ${\cal A}_{1}^{\rm T}$ is
comparable to the simultaneous limit ($|\Re(b_Z)| \lsim  0.013$) 
obtained in our previous analysis~\cite{Biswal:2005fh} using 
unpolarized cross section.

Next we discuss observables $O^{\rm T}_{2}$ and $O^{\rm T}_{3}$ 
constructed using momenta of final state fermions and Higgs boson. 
\begin{figure}[!h]
\begin{center}
\epsfig{file=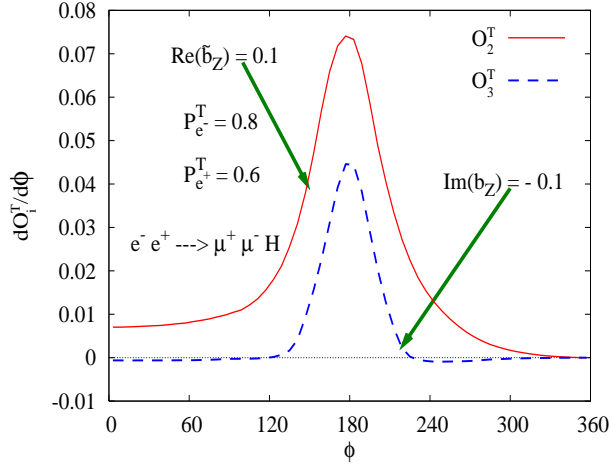,width=8.5cm,height=6.5cm}
\caption{\label{dist_tr}{\em $\phi$-angular dependence of 
integrand of $O^{\rm T}_{2}$ and $O^{\rm T}_{3}$ for final state $\mu$'s 
with 80\%, 60\% transversely polarized $e^-,  e^+$ beams.  
Only the values corresponding to $\Re(\tilde b_Z)$ = 0.1 and 
$\Im(b_Z)$ = - 0.1 respectively are shown, as these are the 
only nonzero contributions. }}
\end{center}
\end{figure}
Fig.~\ref{dist_tr} shows azimuthal angular($\phi$) dependence 
of integrand of $O^{\rm T}_{2}$ and $O^{\rm T}_{3}$ with 
$\mu$'s in the final state for partially transversely polarized 
(${{\cal P}_{e^-}^T}= 0.8$, ${{\cal P}_{e^+}^T} = 0.6$) beams. 
The solid curve is for $\Re(\tilde b_Z)$ = 0.1 
whereas the dotted one is for $\Im(b_Z)$ = - 0.1.  
Here we have integrated over all the phase space 
variables of Eq.~\ref{obs_expt} other than $\phi$, 
the azimuthal angle defined with respect to Higgs boson production plane.  
It is obvious from Fig.~\ref{dist_tr} that the asymmetries 
$O^{\rm T}_{2}$ and $O^{\rm T}_{3}$  are nonzero  and indeed 
can probe specific parts of the anomalous 
$ZZH$ vertex for transversely polarized beams.  

$O^{\rm T}_{2}$  with $R1$-cut for final state muons, electrons 
and for arbitrary transverse beam polarization is given by
\begin{equation}
\begin{array}{rcl}
O^{\rm T}_{2} (R1-\rm{cut}) &=&
\left\{
\begin{array}{lcl}
\displaystyle
\frac{[{{\cal P}_{e^-}^T} {{\cal P}_{e^+}^T}]~
[ 1.19 \ \Re(\tilde b_Z) + 0.02 \ \Im(\tilde b_Z)]}{0.88}
&  & (e^+ e^- H)\\[3.0ex]
\displaystyle
\frac{[{{\cal P}_{e^-}^T} {{\cal P}_{e^+}^T}]~[ 1.19 \ \Re(\tilde b_Z)]}
{0.86} &  & (\mu^+ \mu^- H) \\[3.0ex]
\end{array}
\right.
\end{array}
\label{eq:asym_O7T}
\end{equation}
For the $e^+ e^- H$ channel, $O^{\rm T}_{2}$ receives additional 
contribution from $\Im(\tilde b_Z)$ 
on account of the interference of the $t$-channel diagram
with the absorptive part of the $s$-channel SM one. 
With 80\%(60\%) $e^-$($e^+$) transverse beam polarization
$O^{\rm T}_{2}$ for final state $\mu$'s put a 3 $\sigma$ constraint on
$\Re(\tilde b_Z)$ as: 
\bea
|\Re(\tilde b_Z)| &\leq & 0.22 
\quad {\rm for} \,\, {\cal L} = 500\, {\rm fb}^{-1}.
\label{lim-retbz-m}
\eea

This bound for $\Re(\tilde b_Z)$ using $R1$-cut with 
transverse beam polarization is an improvement by a factor of about $4$  
in comparison to the limit obtained using 
up-down asymmetry ($A_{\rm UD}$) for unpolarized states~\cite{Biswal:2005fh}. 
Since there is no additional 
contribution from the transversely polarized beams to $A_{\rm UD}$, 
the ability of this asymmetry to probe $\Re(\tilde b_Z)$ is not affected 
with the use of transverse beam polarisation. The 
above mentioned improvement with  
transverse beam polarization is possible as $O^{\rm T}_{2}$ is proportional  
to $l_e$ $r_e$ ($\lfsq + \rfsq$) whereas $A_{\rm UD}$ is 
suppressed by ($\lesq - \resq$)($\lfsq - \rfsq$). 
Here $r_f$($l_f$) is the right-(left-)handed coupling of the 
light fermion to the $Z$-boson.  It should be mentioned that  measurements for  
$e^-e^+H$ final state, with $R2$-cut (de-select $Z$-pole events) and
unpolarized beams  can  offer  a better sensitivity to $\Re (\tilde b_Z)$:  
$|\Re(\tilde b_Z)| \leq  0.067$~\cite{Biswal:2005fh}. 
Nevertheless it is useful to have more than one observable 
determining a given coupling and also in complementary kinematic regions.

After imposing all the kinematical cuts of Eqs.~\ref{eq:cuts}, 
along-with $R1$-cut the expression for $O^{\rm T}_{3}$ with  
final state muon for arbitrary transversely polarized beams 
can be written as:

\bea
O^{\rm T}_{3} (R1;~\mu) &=&
\displaystyle
\frac{[{{\cal P}_{e^-}^T} {{\cal P}_{e^+}^T}]~
[- 0.34 \ \Im(b_Z)]}{0.86}
\label{eq:asym_O8T}
\eea
$O^{\rm T}_{3}(R1;~\mu)$ for 80\%(60\%) $e^-$($e^+$)  
beam polarization would lead to a 3 $\sigma$ constraint on 
$\Im(b_Z)$ of the from 

\bea
|\Im(b_Z)| &\leq & 0.77 
\quad {\rm for} \,\, {\cal L} = 500\, {\rm fb}^{-1}.
\label{lim-imbz-m}
\eea

$O^{\rm T}_{3}$ for final state $\mu$ is 
proportional to $l_e$ $r_e$ ($\lfsq - \rfsq$) and   
is suppressed. 
Taking a clue from our earlier analysis for 
measurement of final state $\tau$ polarization along with 
unpolarized/longitudinally polarized beams~\cite{Biswal:2008tg}, we note that  
the sensitivity to $\Im(b_Z)$ can be improved with simultaneous use of 
transverse beam polarization and measurement of  
polarization of the final state $\tau$.  
This reduces the suppression factor present in $O^{\rm T}_{3}$.  
In fact if isolation of events with final state $\tau$'s
in a negative helicity state with an efficiency of 40\% 
is possible, one can constrain $\Im(b_Z)$ to 
$|\Im(b_Z)| \leq  0.24$ using transverse beam 
polarization and $O^{\rm T}_{3}$.  
This in fact compares very favourably 
with the sensitivity limit  $|\Im(b_Z)| < 0.35$ possible using 
 combined polar-azimuthal asymmetry ($A_{\rm comb}$) with $R1$-cut
with unpolarized beams~\cite{Biswal:2005fh}.     

In our previous studies~\cite{Biswal:2005fh,Biswal:2008tg}  
we had observed that the observables constructed 
for unpolarized and longitudinally polarized beams with $R2$-cut
(de-selecting $Z$-pole) can put strong bounds on $\Re(\tilde b_Z)$, 
$\Re(\Delta a_W)$, $\Re(b_W)$ and $\Im(\tilde b_W)$. This is mostly  
due to $t$-channel enhancement in the cross section with 
 $\nu$'s in the final state. There is no such additional gain 
by using transverse beam polarization in this case. This can be 
understood as follows.  
The t-channel squared matrix element (MESQ) never includes 
the spin projection factors ($ 1 + {\gamma}_5 {\not}s_{e^-}$)
and ($ 1 + {\gamma}_5 {\not}s_{e^+}$) in the same trace. Since the
trace of an odd number of $\gamma$-matrices vanishes 
the MESQ for $t$-channel diagram does not have  
any transverse beam polarization dependent term.
Similar observation, which is a consequence 
of electronic chiral symmetry of the SM~\cite{Hikasa:1985qi}, 
has been made for the case of SM $t$-channel diagram~\cite{Hikasa:1984ng}.   
Presence of anomalous $VVH$ vertex does not affect this electronic chiral 
structure; nor does it change 
the $\gamma$-matrix structure in the matrix element. 
Hence there is no additional contribution to the 
$t$-channel matrix element due to transverse beam polarization 
even when anomalous $VVH$ vertex is included. 
For the $s$-channel processes the MESQ includes both the spin
projection factors: ($ 1 + {\gamma}_5 {\not}s_{e^-}$),
($ 1 + {\gamma}_5 {\not}s_{e^+}$) in the same trace and
hence it has nonzero transverse polarization dependent
contribution. It should be noted here that 
the $s$-channel diagram involves two polarized $e^\pm$ at 
the same vertex whereas the $t$-channel diagram has only 
one transversely polarized $e$ at a given vertex.  
Hence additional contributions to the MESQ for 
the $R2$ cut  for the transversely polarized beams
come only  from the interference of the $s$- and $t$-channel diagrams. 
Since imposition of $R2$-cut reduces the $s$-channel contribution, 
the observables: $O^{\rm T}_{1}$, $O^{\rm T}_{2}$ and $O^{\rm T}_{3}$ 
with $R2$-cut are less sensitive than those with $R1$-cut.

%
\subsection{\label{sec::wwh}Anomalous $WWH$ Couplings}
As discussed before 
observable $O^{\rm T}_{1}$ can probe some parts of $CP$-even 
anomalous $VVH$ couplings. 
However, note here that 
transverse beam polarization does not affect the 
squared matrix element of the $t$-channel $WW$ fusion diagram 
which includes the anomalous $WWH$ couplings. As a result 
terms proportional to anomalous $WWH$ couplings in $O^{\rm T}_{1}$ 
receive contribution only from the interference of
$t$-channel diagram with the $s$-channel SM part. Hence $O^{\rm T}_{1}$  
is not expected to put stronger bounds on 
$CP$- and $\tilde T$-even anomalous $WWH$ couplings 
compared to the bounds obtained for unpolarized beams\cite{Biswal:2005fh}.  
As a result, we have not explored numerically the reach in 
sensitivity of this observable with $\nu$'s in the final state.  

\section{\label{sec:conclude}Summary}
We have considered the $VVH$ ($V = Z/W$) anomalous
couplings to be complex and investigated use of transverse beam 
polarization in probing these couplings. This is an extension of 
our earlier work~\cite{Biswal:2005fh,Biswal:2008tg} 
assessing the use of unpolarized or longitudinally 
polarized beams to probe these anomalous couplings.
We have constructed observables with definite discrete 
transformation ($CP$, $\tilde T$) properties that are 
sensitive to a single anomalous coupling
and hence can probe the corresponding operator in the effective Lagrangian
with the same $CP$ and $\tilde T$ properties. 
We have imposed various kinematical cuts on different
final state particles to reduce backgrounds and considered 
the events wherein the $H$ decays into 
a $b \bar b$ pair with branching fraction $\sim 0.68$, 
with a $b$-tagging efficiency of 70\%,  
same as in our earlier 
analyses~\cite{Biswal:2005fh,Biswal:2008tg}. 

Use of transverse beam polarization to constrain 
anomalous $ZZH$ and $Z \gamma H$ interactions 
has been studied previously in Refs.~\cite{Hagiwara:1993sw,Rindani:2009pb}. 
In Ref.~\cite{Hagiwara:1993sw} the anomalous $ZZH$ couplings were 
assumed to be real and it was observed that 
polarization of beams does not provide any new information
about the anomalous $ZZH$ couplings.  
Ref.~\cite{Rindani:2009pb} had considered 
these couplings to be complex, but they had investigated use of  
$Z$-angular distribution in the process $e^+e^- \to ZH$, which cannot 
probe $\Im(b_Z)$ and $\Re(\tilde b_Z)$. 
In our analysis, we find that it is possible to have independent  
probes to constrain $\Re(\tilde b_Z)$, $\Im(b_Z)$,  
$\Re(b_Z)$ and $\Delta a_Z$ 
through a study of the process $e^+ e^- \to f \bar f H$  
using transverse beam polarization. 
Ref.~\cite{Rao:2006hn} contains an analysis of 
the four point $e^-e^+ZH$ coupling with transverse beam 
polarization in the Higgs-strahlung
($s$-channel) process $e^+e^- \rightarrow ZH  \rightarrow l^+ l^- H$. 
The general $ZZH$ anomalous couplings can be put into a 
one to one correspondence with the four point $e^-e^+ZH$ coupling,  
but only with momentum dependent factors.  
Hence it is not possible to translate the results of 
Ref.~\cite{Rao:2006hn} directly to the case of  
$VVH$ interactions investigated here.  

Use of transverse beam polarization allows us to construct completely 
independent probes of both the $CP$- and $\tilde T$-even 
couplings, $\Re(b_Z)$ and $\Delta a_Z$. This was not possible with 
unpolarized states~\cite{Biswal:2005fh}.
In Ref.~\cite{Biswal:2008tg} we had used
numerical combinations of various linearly polarized cross
sections to reduce the contamination from $\Re(b_Z)$
in probing $\Delta a_Z$ and vice versa, wherein
the numerical coefficients depend on cuts employed as well as
on the exact size of SM contribution etc.
On the other hand  
use of transverse beam polarization
can probe both these couplings individually
which lead to independent determination of all the anomalous
parts of the $ZZH$ vertex. 
Further, it helps to improve 
on the sensitivity of $\Re(\tilde b_Z)$ by a factor of about $4$ 
for s-channel final state muons. It is possible 
to get rid of the suppression factor 
($\lfsq - \rfsq$) in $O^{\rm T}_{3}$ 
by isolation of events with a final state $\tau$ in definite helicity 
state, which can then lead to increased sensitivity.  

To summarize, use of transverse beam polarization helps to 
construct independent probes of both the $CP$- and $\tilde T$-even
$ZZH$ couplings. It can also 
improve the sensitivity to probe one of the $CP$-odd 
anomalous part of the $ZZH$ vertex by a factor of up to $4$-$5$ 
for muons in the final state. Use  
of transverse beam polarization  
along with measurement of final state $\tau$ polarization can 
increase the sensitivity to $\Im(b_Z)$ also by 
30\% compared to the unpolarized case. 
\vspace{-0.5cm}
\section*{Acknowledgments} 
We thank S. D. Rindani for useful discussions and a careful reading of 
the manuscript. R. M. G. wishes to acknowledge the 
Department of Science and Technology, India for financial support 
under Grant No. SR/S2/JCB-64/2007. 

\end{document}